\documentstyle[mprocl]{article}

\input epsf
\def\be{\begin{equation}}
\def\ee{\end{equation}}
\def\bea{\begin{eqnarray}}
\def\eea{\end{eqnarray}}

\begin{document}
\begin{center}
\title{\bf Extra Dimensions and Black Hole
Production\footnote{Invited Talk given at the International
Conference: {\it New Trends in High Energy Physics}, Yalta
(September 23-28, 2001).}}
\end{center}

\author{C. Pagliarone}

\address{I.N.F.N. of Pisa\\
via Livornese, 1291 - 56010 S. Piero a Grado (PI) - ITALY\\
E-mail: pagliarone@fnal.gov}

\vspace*{0.3cm} \maketitle \abstracts{If nature realizes $TeV$ scale
gravity, we are entering a very exciting period in which
we could be able to  address, experimentally, some questions on
quantum gravity, strings, branes and other exotic aspects of the
fundamental theory of gravity. This article reviews recent
development in models with Large Extra Dimensions and Black hole
production at future colliders. Experimental results from current
experiments as well as the expectation for the future colliders
are summarized.}

\section{Introduction}

The Standard Model (SM) has proved to be enormously successful in
providing a description of particle physics up to energy scales of
several hundred GeV as probed by current experiments~\cite{SM-exp}.
In the SM, however, one assumes that effects of gravity can be neglected,
because the scale where such effects become large is the
Planck Scale.
The question of why the $4-$dimensional Planck Scale,
$G^{-1/2}_{Pl} \sim 10^{19}$ GeV, is much larger than the
electroweak (EWK) scale, $G^{-1/2}_{{\mathcal{F}}} \sim 10^{2}$ GeV, is an
outstanding problem in contemporary physics. The {\it hierarchy
problem} is in the essence the difficulty to explain the large
disparity between these two numbers~\cite{Hyerarchy}.

\noindent Motivated in part by naturalness issues, numerous
scenarios have emerged recently, that address the hierarchy
problem within the
context of the old idea that some part of the physical world (i.e.
the SM-world) is confined to a brane in a higher dimensional
space~\cite{kaluza-klein}. Although supergravity theories were
formulated up to $11$ dimensions and Superstring theories in $10$
dimensions were known since the 70's, the idea to extend this
extra spatial dimension paradigm ($ESD$) to other contexts,
received a new impulse only recently~\cite{HDD1}.
As we don't experience in our
everyday world, more then $3$ spatial dimensions, we have to assume
that any possible $ESD$ is hidden.

\noindent
There is a simple and elegant way to hide possible extra spatial
dimensions: the {\it compactification}. The result is achieved by
assuming, for example, that the extra dimensions form, at each point
of the $4-$dimensional space, a torus of volume
$(2\pi)^{{\mathcal{D}}}R_{1}R_{2}...R_{{\mathcal{D}}}$.
In this way it is possible to allow the gravity to live in the ${\mathcal{D}}$ large
extra dimensions, the {\it bulk}, while the SM fields
will lie on a 3-${\mathcal{D}}$ surface, the {\it brane}.

\noindent In presence of a compactfied extra spatial dimension
$y$, a field $\phi(x_{\mu},y)$ of mass $m_{0}$ is periodic over
$y$ and can be Fourier developed as follow:
\begin{equation}
\phi(x_{\mu},y) =\sum_{k=-\infty}^{+\infty}
e^{\frac{iky}{R}}\phi^{(k)}(x_{\mu})
\end{equation}
where $R$ is the radius of the compact $ESD$. The $4{\mathcal{D}}$
terms $\phi^{(k)}(x_{\mu})$ are the Kaluza-Klein states ($KK$),
also called {\it modes} or {\it excitations}. The mass of each $KK$ mode is
then expressed by the formula:
\begin{equation}
m^{2}_{k}= m_{0}^{2} + k^{2}/R^{2}
\end{equation}

\noindent If $M_{\mathcal{F}}$ is the actual fundamental scale of
gravitational interactions and if $V_{extra}$
is the volume of the extra dimensional ${\mathcal{D}}-$fold, then,
by Gauss's law, at distances larger than the inverse mass of the lightest
$KK$ mode in the theory, the gravitational force will follow an
inverse square law with an effective coupling of:
\begin{equation}
M_{Pl}^{-2} = M_{{\mathcal{F}}}^{-({\mathcal{D}}+2)}V_{extra}^{-1}
\end{equation}
We see indeed that gravity becomes strong in the full
$4+{\mathcal{D}}-$dimensional space at a scale $M_{{\mathcal{F}}}$
of few $TeV$ which is far below the conventional Planck Scale
($M_{Pl}^{2} = M_{{\mathcal{F}}}^{{\mathcal{D}}+2}V_{extra}$).

\noindent At the present there are many models which assume the
existence of extra spatial dimensions predicting the appearance of
new physics signatures that can be probed at energy scale above
$1$ $TeV$. Most of the models fall into one
of the three following classes:

\subsection{The Large Extra Dimensions Scenario}

The large extra dimension scenario ($LED$) started with the
works of Arkani-Hamed, Dimopoulos and Dvali ($ADD$)~\cite{HDD1,HDD2}.
In this model the SM particles live on a $3+1-$dimensional space
($3-$brane) while the gravity is free to propagate in
higher-dimensional space, extra dimensions. This model predicts
essentially the emission and exchange of large Kaluza-Klein towers
of gravitons that are finely-spaced in mass.
The $ADD$ Model was first proposed to solve the hierarchy problem
by requiring the compactified dimensions to be of very large size.

\subsection{KK gauge bosons}

A second possibility comes from all those models where
the extra spatial dimensions are of $TeV$ scale size.
In these class of models there are $KK$ excitations of the SM gauge fields
with masses of the order a $TeV$ which can show up in collider experiments
as resonances.

\subsection{Warped Extra Spatial dimension}

Another approach for extra spatial dimensions has been
proposed by L. Randall and R. Sundrum
($RS$ models or $WED$)~\cite{Randall-Sundrum}.
In this scenario two $4{\mathcal{D}}$ branes with tension $V$ and $V^{'}$
are situated in the position, $y=0$ and $y=\pi r_{c}$, of a
$5{\mathcal{D}}$ bulk with cosmological constant $\lambda$,
where the gravitation lives.
One of the interesting consequence of this assumptions is the fact that
the fundamental mass scale on the brane at $y=0$ is then red-shifted
by a factor $e^{-2k|y|}$ ({\it warp factor}) on the other brane
at  $y=\pi r_{c}$. \,In this way the EWK \,scale \,\,(${\mathcal{O}}(1\,\,TeV)$)
can be produced \,\,from

%
%
\begin{table}[t!]
\medskip
\centering
\begin{tabular}{lcccc}
\hline\hline
Observation \& Object&
\multicolumn{2}{c}{$f_{\rm KK}^{\rm max}$}&
\multicolumn{2}{c}{$M^{\rm min}$ [TeV]}\\
&$n=2$&$n=3$&$n=2$&$n=3$\\
\hline\hline
\noalign{\smallskip}
\multicolumn{4}{l}{Neutrino signal}\\
\quad SN~1987A~\protect\cite{Cullen:1999hc,Barger:1999jf,%
Hanhart:2001er,Hanhart:2001fx}
&0.5&0.5&31&2.75\\
\hline
\multicolumn{4}{l}{EGRET $\gamma$-ray limits}\\
\quad Cosmic SNe~\protect\cite{Hannestad:2001jv}
&$0.5\times10^{-2}$&$0.5\times10^{-2}$&84&7\\
\quad Cas A&$0.8\times10^{-2}$&$0.3\times10^{-2}$&87&8\\
\quad PSR J0953+0755&$2.2\times10^{-5}$&$0.9\times10^{-5}$&360&22\\
\quad RX J185635--3754&$0.5\times10^{-5}$&$1.8\times10^{-6}$&540&31\\
\hline
\multicolumn{4}{l}{Neutron star excess heat}\\
\quad PSR J0953+0755& $0.5 \times 10^{-7}$ & $0.5 \times 10^{-7}$
& 1680 & 60 \\
\hline
\multicolumn{4}{l}{GLAST $\gamma$-ray sensitivity}\\
\quad RX J185635--3754&$1\times10^{-7}$&$0.5 \times 10^{-7}$&1300&60\\
\hline \hline
\end{tabular}
\caption{\label{SNlimits}Constraints on the fundamental
scale $M_{\mathcal{F}}$ coming from SN and neutron stars.}
\end{table}

\noindent
the Planck Scale.
In the $RS$ model the $4{\mathcal{D}}$ Planck mass is then expressed
by the formula:
\begin{equation}
M^{2}_{Pl} = \frac{M^{3}_{5}}{k}[1-e^{-2kr_{c}\pi}]
\end{equation}

\section{Experimental Astrophysical Constrains on LED}

Various astrophysical and cosmological processes can be used
to set limits on the model with extra dimensions. The
phenomenology of the Supernova SN1987A places strong constrains on
the energy loss mechanism, allowing to derive a bound on the
fundamental Planck Scale. In fact if $LED$ exist, then the usual $4D$
graviton is complemented by a tower of Kaluza-Klein states,
corresponding to the new available phase space in the bulk. These
$KK$ gravitons would be emitted from the Supernova core after
collapse, complete with neutrino cooling, and shorten the
observable signal. This argument has led to the following bounds
that require that the fundamental scale be as high as $M \ge 84$
$TeV$ assuming $N=2$ or $M \ge 7$ $TeV$ for $N=3$~\cite{cosmos}.
The limits on the fundamental scale $M_{\mathcal{F}}$ coming from
SN and neutron star data are summarized in Table~\ref{SNlimits}.

\noindent
Other effects are the possible distorsion of the cosmic diffuse
gamma background because of the new accessible production
mechanism: $G_{KK} \rightarrow \gamma \gamma$ decays. Using the
COMPTEL data in the $E_{\gamma}$ range $[800$ $KeV - 30$ $MeV]$,
Hall and Smith set a lower limit on effective Planck scale
$M_{\mathcal{F}}$ as function of the number of $LED$
($\mathcal{D}$): $M_{\mathcal{F}}> 100$ $TeV$ (${\mathcal{D}}=2$)
and $M_{\mathcal{F}}> 5$ $TeV$
(${\mathcal{D}}=3$)~\cite{Hall-Smith}.
Analogous analysis performed by Hannestad and Raffelt on EGRET
data ($E_{\gamma} \sim [3-200]$ $MeV$) set a limit on $M_{\mathcal{F}}$ as
function of the number of $LED$: $M_{\mathcal{F}}> 84$ $TeV$
(${\mathcal{D}}=2$) and  $M_{\mathcal{F}}> 7$ $TeV$
(${\mathcal{D}}=3$)~\cite{Hannestad:2001jv}.

%
%
\newpage
\begin{figure}[t!]
\hspace{0.2cm}
 \begin{minipage}{8.0in}
  \epsfxsize2.0in
  \vspace{-2.7cm}
  \hspace*{-0.4cm}\epsffile{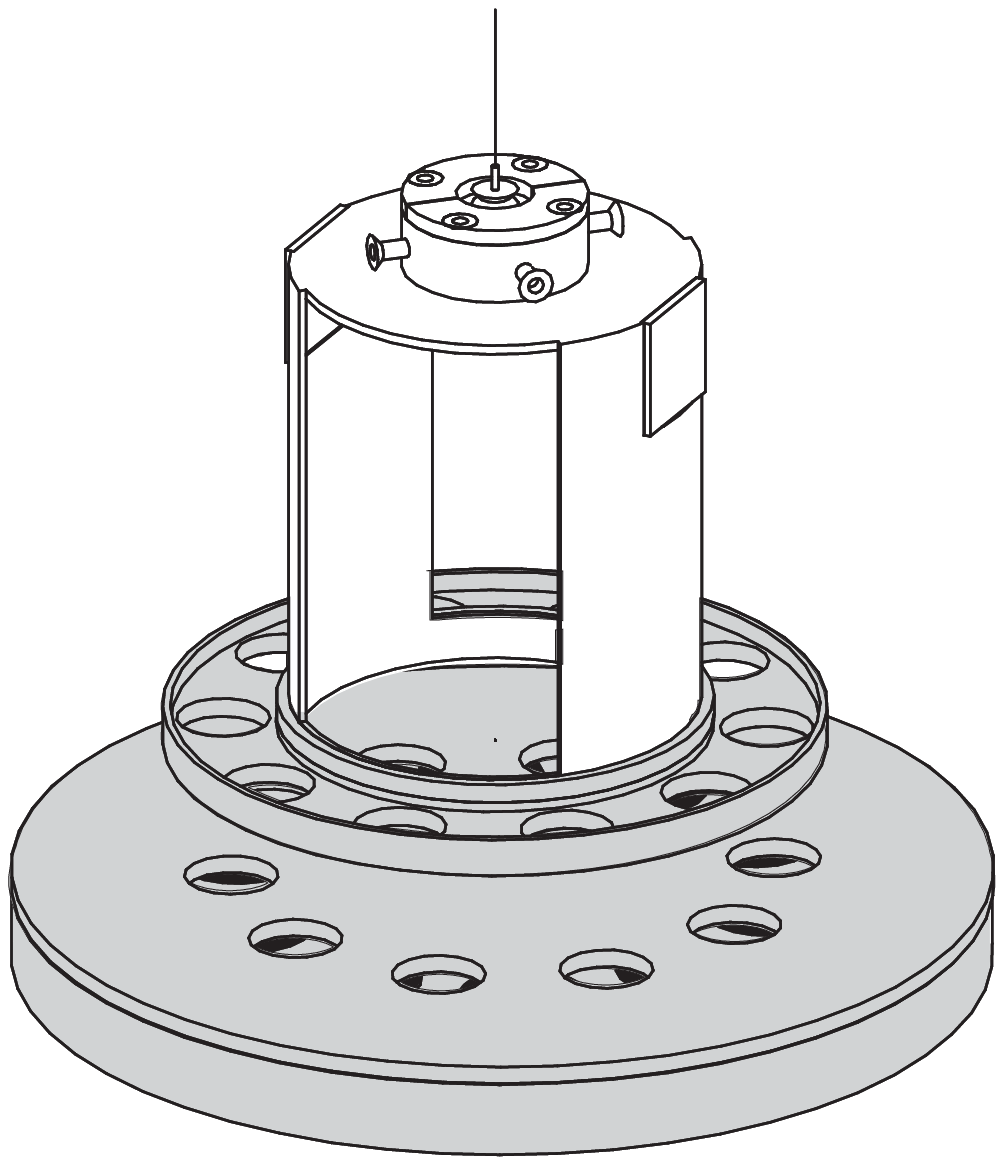}
  \epsfxsize4.1in
  \hspace*{-0.4cm}\epsffile{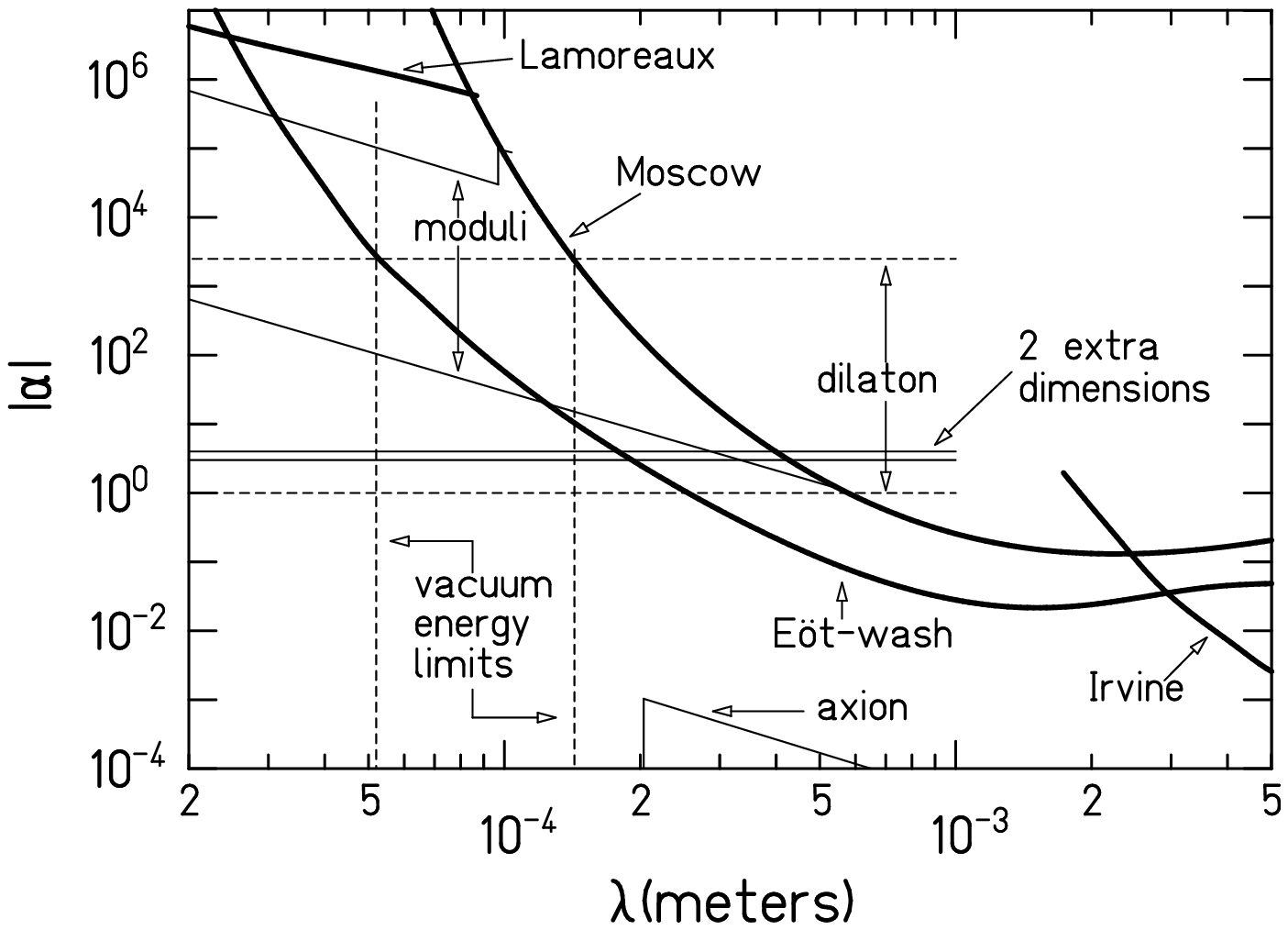}
 \end{minipage}\hfill
\renewcommand{\baselinestretch}{1.0}
\caption{\it  a) Scale drawing of the torsion pendulum and
rotating attractor; the active components are shaded and the
distance between these two parts have been enhanced for clarity.
b) 95\% C.L. upper limits on the violation of the $1/R^{2}$ Newton
Law as described in equation~\ref{newton-law}. The regions
excluded by previous works lies above the lines labeled Irvine,
Moscow and Lamoreaux, respectively.}
 \label{direct-exp}
 \end{figure}

\section{Direct Gravity Experiments}

Until the middle of 1970 years a number of Cavendish-type
experiments, searching also for the so called ``fifth forth'',
have been performed in order to test possible deviation of
Newton's gravitational interaction. This kind of experiments
tested with good precision the Newton's law for a mass separation
greater than $1$ $cm$. As a matter of fact, in such experiments, the
sensitivity vanishes quickly for distances below $1$ $mm$ because
of the effects due to the Casimir and Van der Waals forces.
Short-range regime of gravitation was not explored further because
it was generally assumed that non-relativistic gravity obey an
$1/R$ law for all $R >> R_{Pl} = \sqrt{G\hbar/c^{3}}$. The
higher-dimensional theories described above suggest nowadays that
new effect may show up at short distances. The string theory in
particular predict scalar particles as {\it moduli} or {\it
dilatons} that generate Yukawa interactions which could modify the
Newton Gravitational Potential:
\begin{equation}
V(R) = -G\frac{ m_{1} m_{2} }{R}\left( 1 + \alpha e^{ -R/\lambda }
\right)
\label{newton-law}
\end{equation}
In the simplest scenario, with a number of extra
spatial dimensions equal to $2$,
we obtain the following values for
$\lambda$ and $\alpha$: $\lambda= R^{*}$, $\alpha=$ $3$ for a
compactification on an 2-sphere and $\lambda= R^{*}$, $\alpha=$
$4$ for a compactification on an 2-torus, where $R^{*}$ is the
extra dimension radii:
\begin{equation}
R^{*} = \frac{\hbar c}{M^{*}c^{2}} \left( \frac{M_{Pl}}{M^{*}} \right)
\end{equation}

\noindent and $M^{*}$ is the unification scale (usually taken as $M_{SM}$).
Recently sub-millimeter test of the gravitational
inverse-square law have been performed by C.D. Hoyle {\it et
al.}~\cite{Hoyle:2001cv} using a $10$-fold symmetric torsion
pendulum and a rotating $10$-fold symmetric attractor. A schematic
description of the experimental setup is given in
Fig.~\ref{direct-exp}.a.
The results obtained, if interpreted in
the simplest unification scenario with $2$ equal large extra
dimensions, \,\,imply \,a \,unification \,scale of \,\,$M^{*} \geq 3.5$ $TeV$
(see Fig.~\ref{direct-exp}.b).

%
%
\newpage
\begin{figure}[t!]
\hspace{-1.cm}
 \begin{minipage}{8.0in}
  \epsfxsize3.3in
  \vspace{-2.8cm}
  \hspace*{-0.9cm}\epsffile{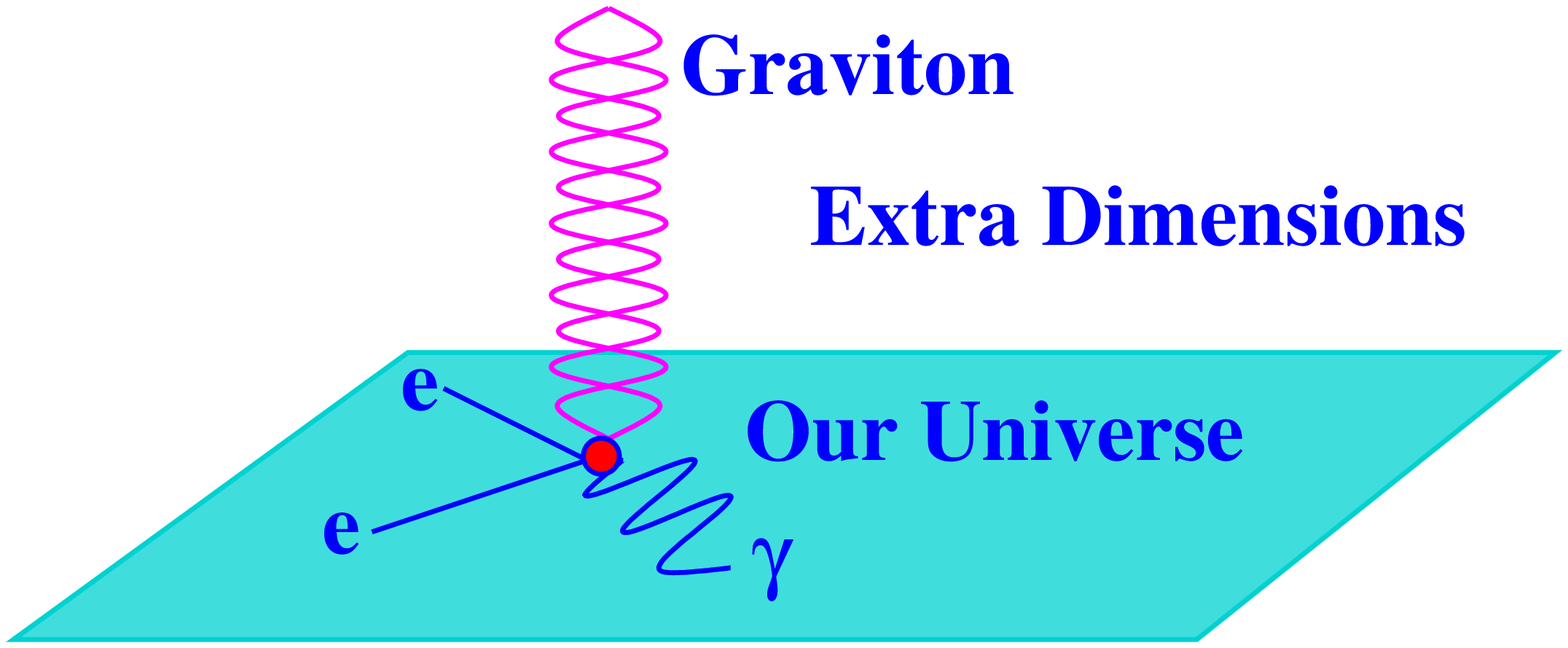}
  \epsfxsize3.3in
  \hspace*{-0.6cm}\epsffile{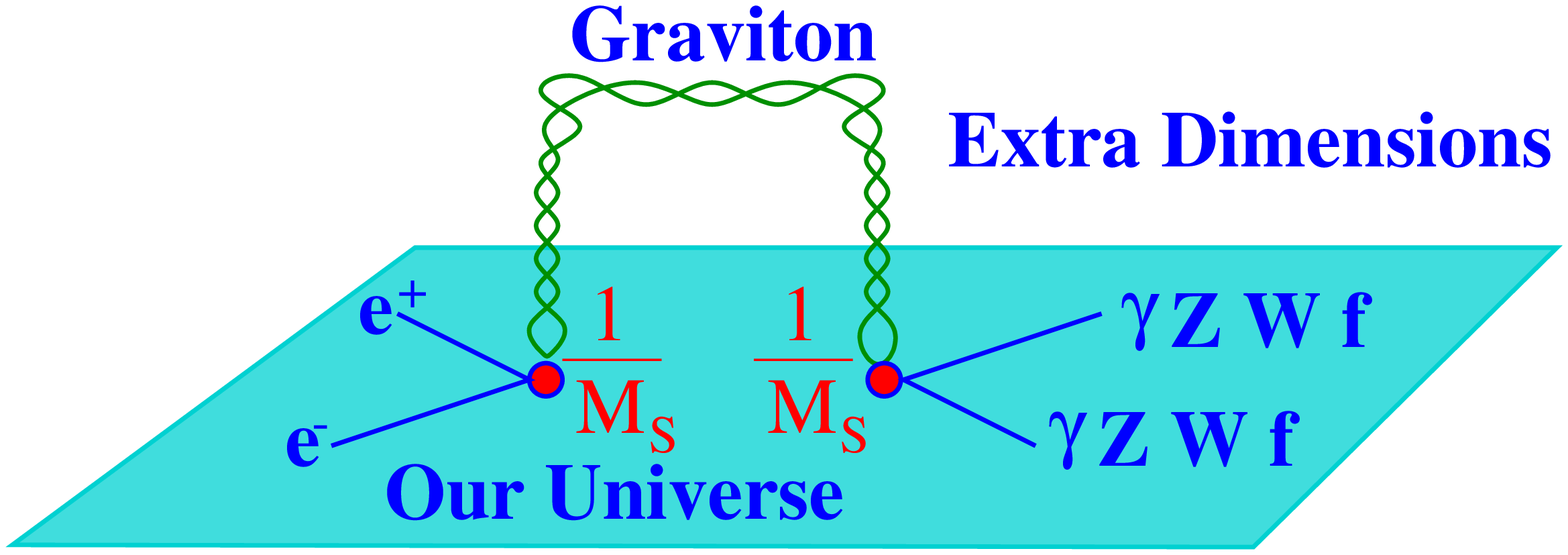}
 \end{minipage}\hfill
\hspace{-2.3cm}
 \caption{\it Direct graviton emission diagram contributing
to the process $e^{+}e^{-} \rightarrow \gamma G$ (left);
Virtual graviton exchange diagrams (right).}
 \label{h1h2}
 \end{figure}
\section{LED at present collider Experiments}

Most of the searches performed until now for large extra
dimensions have been done assuming the $ADD$ phenomenology. We
summarize here signatures and results of such searches.

\noindent
As Kaluza-Klein gravitons couple to the momentum tensor,
they therefore contribute to most of the SM processes.
Depending on whether the  $G_{KK}$ leaves our world or
remains virtual, the collider signatures change.
For graviton that propagate in the bulk, in particular,
from the point of view of our $3+1$ space-time,
energy and momentum are not conserved in the $G_{KK}$ emission.
Gravitons, on the other hand, interacting weakly with detectors,
escape detection causing a typical missing transverse
energy ($\not\!\!\!E_{\rm T}$) signal.
We will summarize in the next two paragraphs the different classes
of signatures that correspond experimentally to the real and to
virtual graviton emissions.

\subsection{Real Graviton Emission}

Direct emission of gravitons leads to the presence of missing
energy in the final state. At $e^{+}e^{-}$ colliders the best
signal is the associated production of gravitons with a Z boson, a
photon, or a fermion pair. In the hadron colliders the prominent
and most studied signature is the production of one jet (monojet)
associated with large transverse energy in the final state.

\noindent The effects of direct graviton production, including
single photon or Z's production, have been studied at
LEPII~\cite{gammaG,gammaZ}. The
following final states have been considered in the analysis:
$\gamma +$$\not\!\!\!E_{\rm T}$, $Z(\rightarrow
jj)+$$\not\!\!\!E_{\rm T}$ (see Fig.~\ref{h1h2}).
With no excess apparent beyond the SM
expectations, a lower limit on the graviton contribution have been
calculated at 95\% Confidence Level. The limits, expressed in
terms of effective Planck Scale, are summerized in
Table~\ref{LEP-limits}.

\noindent The CDF and D\O \,\, Collaborations at the Fermilab
Tevatron Collider also looked for direct graviton emission. From
an experimental point of view the mono-jet plus
missing$-\not\!\!\!E_{\rm T}$ signature is quite complex to study
because of the large instrumental background from jet
mismeasurement and the presence of cosmic rays background.
Results on this channel will be available soon.

\subsection{Virtual Graviton emission}

The virtual exchange of graviton towers either leads to
modifications in SM cross sections and asymmetries or to new
processes not allowed in the SM at the tree level. Collider
signatures with virtual exchanges of $KK-$gravitons are several
and include diphoton, diboson and fermion-pair production. In the
case of virtual $G_{KK}$
%
%
\newpage
\renewcommand{\baselinestretch}{3.40}
\begin{table}[h!]
\begin{tabular}{|l|c|c|c|c|c|c|}
\hline\hline
& \multicolumn{2}{c|}{\bf N=2}&\multicolumn{2}{c|}{\bf N=4}& \multicolumn{2}{c|}{\bf N=6}\\
\cline{2-7}
& $M_{D}(TeV)$ & $R(cm)$ &  $M_{D}(TeV)$ & $R(cm)$ &  $M_{D}(TeV)$ & $R(cm)$ \\
\hline\hline
ALEPH & 1.28 & 2.9 $\cdot 10^{-2}$ & 0.78 & 1.4 $\cdot 10^{-9}$ & 0.57 & 5.6 $\cdot 10^{-12}$ \\
DELPHI & 1.38 & 2.5 $\cdot 10^{-2}$ & 0.84 & 1.3 $\cdot 10^{-9}$ & 0.58 & 5.4 $\cdot 10^{-12}$ \\
L3 & 1.45 &  2.3 $\cdot 10^{-2}$ & 0.87 &  1.2 $\cdot 10^{-9}$ & 0.61 & 5.2 $\cdot 10^{-12}$ \\
\hline\hline
\end{tabular}
\renewcommand{\baselinestretch}{1.0}
\caption{\it 95\% C.L. lower limits on the gravitational scale $M_{D}$ and on the size of
extra dimensions $R$, derived from the direct graviton searches in the
$e^{+}e^{-} \rightarrow \gamma G$ channel}
\label{LEP-limits}
\end{table}
\noindent
emission, gravitons  lead to apparent
violation of $4$-momentum as well as of the angular momentum.

\noindent
There are several processes that can be studied at lepton
colliders~\cite{LEP-virtual}:
$e^{+}e^{-}\rightarrow \gamma \gamma$, $e^{+}e^{-}$,
$W^{+}W^{-}$, $Z^{0}Z^{0}$ (see Fig.~\ref{h1h2}).
Among many fermion and boson-pair
final states studied at LEP, the most sensitive channels involve
Bhabha scattering ($e^{+}e^{-} \rightarrow e^{+}e^{-}$) and the
photon-pair production ($e^{+}e^{-} \rightarrow \gamma \gamma$).
The lower limits at 95\% C.L. on $M_{\mathcal{F}}$ obtained using
the $e^{+}e^{-}$ and $\gamma \gamma$ channel are given in
Table~\ref{LEP-1} and in Table~\ref{LEP-2}.
The combined limit, as obtained  by assuming log likelihood curves
from different channels, is $M_{\mathcal{F}} >$ $1.03$ $TeV$ for
$\lambda=+1$ and $M_{\mathcal{F}} >$ $1.17$ $TeV$ for
$\lambda=-1$.
\renewcommand{\baselinestretch}{3.40}
\begin{table}[h!]
\centering
\begin{tabular}{|c|c|c|c|c|}
\hline\hline $\lambda$    & L3   & OPAL \\ \hline $\lambda=+1$ &
1.06 & 1.00 \\ $\lambda=-1$ & 0.98 & 1.15 \\ \hline\hline
\end{tabular}
\renewcommand{\baselinestretch}{1.0}

\caption{\it 95\% C.L. lower limits on $M_{\mathcal{F}}$ obtained
by L3 and OPAL using the Bhabha scattering process: $e^{+}e^{-}
\rightarrow e^{+}e^{-}$; results are shown in Hewett notation.}
\label{LEP-1}
\end{table}

\noindent
The impact of virtual gravitons in hadron collider
experiments can be observed in processes such as: $q \bar{q}
\rightarrow G \rightarrow \gamma \gamma$ or $g g \rightarrow G
\rightarrow e^{+}e^{-}$ where the ADD model introduces production
mechanism that can increase the cross-section of diphoton and
dielectron production at high invariant mass over the SM. The
diphoton and dielectron cross-section considering the $LED$
contributions take the form~\cite{HDD2}:
\begin{equation}
\frac{d^{2}\sigma_{Tot}}{d cos\theta^{*}\,\,  dM} =
\frac{d^{2}\sigma_{SM}}{d cos\theta^{*}\,\,  dM} \,\,+\,\,
\frac{a(n)}{M^{4}_{\mathcal{F}}} \,\,F_{1}(cos \theta^{*}, M) \,\,+\,\,
\frac{b(n)}{M^{8}_{\mathcal{F}}} \,\,F_{2}(cos \theta^{*}, M)
\label{ccc}
\end{equation}
where $cos\theta^{*}$ is the scattering angle of the photon or
electron in the center of mass frame of the incoming parton.
The first term in the expression~\ref{ccc} is the pure SM contribution
to the cross section; the second and the third part are
the interference term and the direct $G_{KK}$ contribution.
The characteristic signatures for contributions from virtual $G_{KK}$
correspond to the formation of massive systems abnormally beyond
the SM expectations. Figure~\ref{D0fig} shows a comparison of the
two-dimensional distributions in di-EM mass and $|cos\theta^{*}|$
for data, SM background processes and for background plus $LED$
contribution as obtained by D\O \,\,Collaboration.
With no excess apparent beyond expectations of the
SM, D0 proceeds to calculate a lower limit on the graviton
contribution to the di-EM cross section. Hence there are three
main formulation on the effective lagrangian the Giudice, Rattazzi
and Wells (GRW)~\cite{GRW}, the Han, Lykken and Zhang (HLZ)~\cite{HLZ}
notation and the Hewett~\cite{Hewett} one the limits in
%
%
\renewcommand{\baselinestretch}{3.40}
\begin{table}[h!]
\centering
\begin{tabular}{|c|c|c|c|c|}
\hline\hline $\lambda$    & ALEPH & DELPHI & L3   & OPAL \\ \hline
$\lambda=+1$ & 0.81  & 0.82   & 0.83 & 0.83 \\ $\lambda=-1$ & 0.82
& 0.91   & 0.99 & 0.89 \\ \hline\hline
\end{tabular}
\renewcommand{\baselinestretch}{1.0}

\caption{\it 95\% C.L. lower limits on $M_{\mathcal{F}}$ obtained
using $\gamma \gamma$ channel; the results are given in the Hewett
formulation.} \label{LEP-2}
\end{table}

\noindent
the Table~\ref{D0tab} are translated in
all this notations as well as function of the number of extra dimensions
for the HLZ approach~\cite{D0}.

\noindent
The CDF Collaboration performed a similar search using the
dielectron and diphoton channels. The combined results expressed
in the Hewett convention for $\lambda=+1$ is
$M_{\mathcal{F}}>$ $0.814$ $TeV$ and for
$\lambda=-1$ are
$M_{\mathcal{F}}>$ $0.9$ $TeV$~\cite{CDFnote}.

\noindent
More studies are forthcoming from CDF and D\O on real
graviton emission (mono-jet events), as well as on
virtual graviton exchange, which by the end of Run-II,
should be sensitive to scales of $2-3$ $TeV$.
Beyond that lies LHC which sensitivity should cover a mass region
up to $M_{\mathcal{F}} \sim 10$ $TeV$~\cite{cheung}.
The sensitivity of Tevatron and LHC experiments on
$M_{\mathcal{F}}$ for ${\mathcal{D}}=$ $4$ is summarized
in Table~\ref{future}.

\section{LED Searches at Future Colliders}

Different signatures for several extra dimensional models
have been studied recently in order to understand which results will be possible
to obtain at future lepton colliders such as CLIC
or hadron colliders as the Very Large Hadron Collider (VLHC)
(see Table~\ref{future}).

\noindent
CLIC is an $e^{+}e^{-}$ collider with an expected
center of mass of energy ranging between $3$ and $5$ $TeV$.
With and integrated luminosity of $1$ $ab^{-1}$
for many of the extra dimensional models the
experimental reach is found to
be in the interval of $\sim 15-80$ $TeV$~\cite{Rizzo-1}.

\noindent
The VLHC is a hadron collider thought to be build in 2
steps: a stage I in which the avaible
center of mass energy will be $40$ $TeV$
and a stage II when, by using new magnets, the center of mass
energy could reach a value of $175$ or $200$ $TeV$.
In particular, as have been shown by T.~Rizzo~\cite{Rizzo-2}
a $200$ $TeV$ VLHC with $1$ $ab^{-1}$ of an integrated
luminosity will be able to observe the first $RS$ $KK$
excitation for masses as large as $15-30$ $TeV$
and values of $M_{\mathcal{F}} \sim 50-60$ $TeV$, in the ADD model,
will be directly probed in the Drell-Yan processes.

\section{Black Hole Production at Hadron Colliders}

Since long ago, black holes ($BH$) have been objects of interest in theoretical
physics and astrophysics, and more recently also in experimental
high energy physics. As a matter of fact, $BH$ can be produced in a
particle collisions if the center of mass

\newpage
\begin{figure}[t!]
\vspace{1.2cm}
 \begin{minipage}{7.0in}
  \epsfxsize3.0in
  \hspace*{-1.6cm}\epsffile{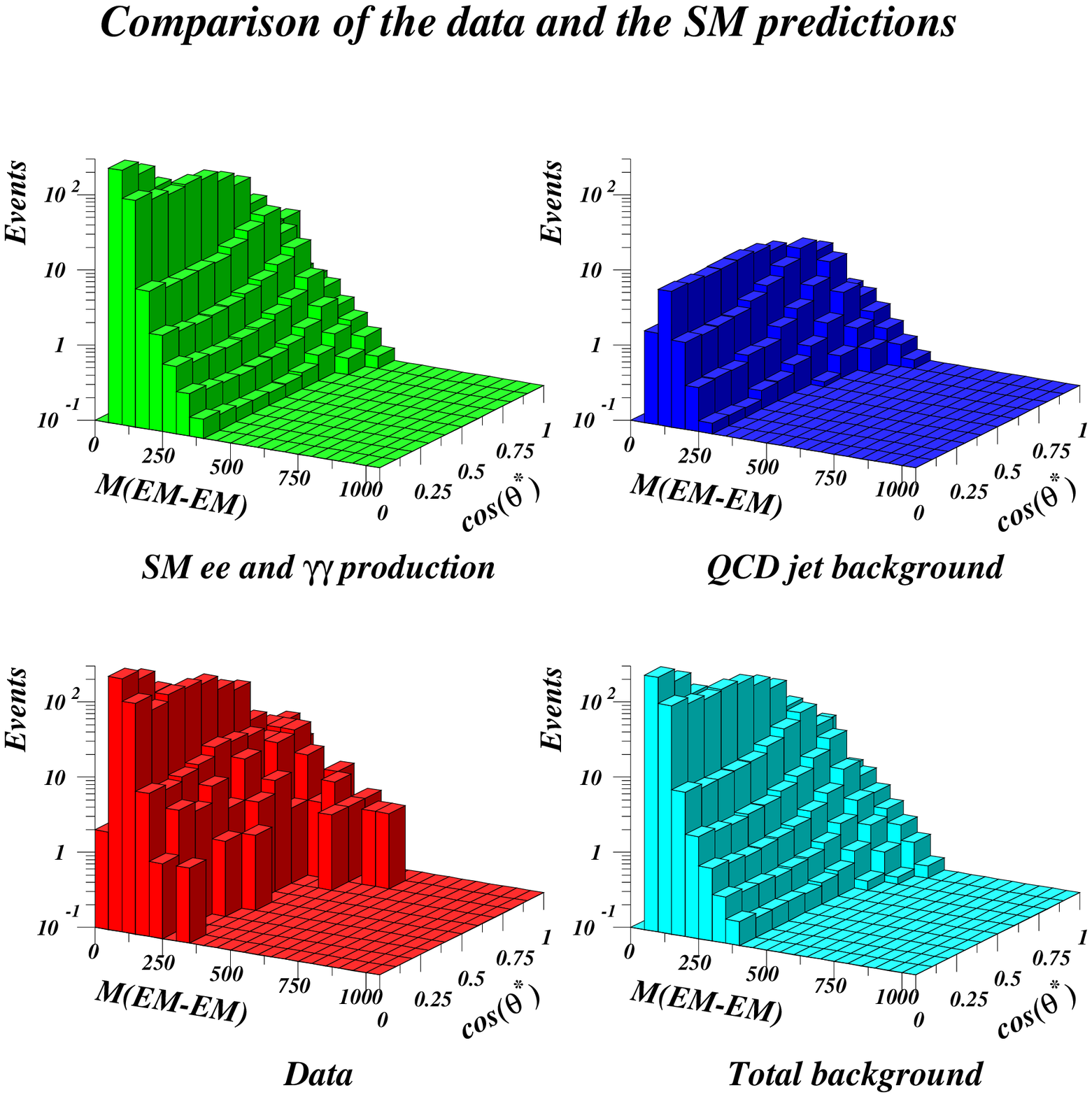}
  \epsfxsize3.0in
  \hspace*{-0.2cm}\epsffile{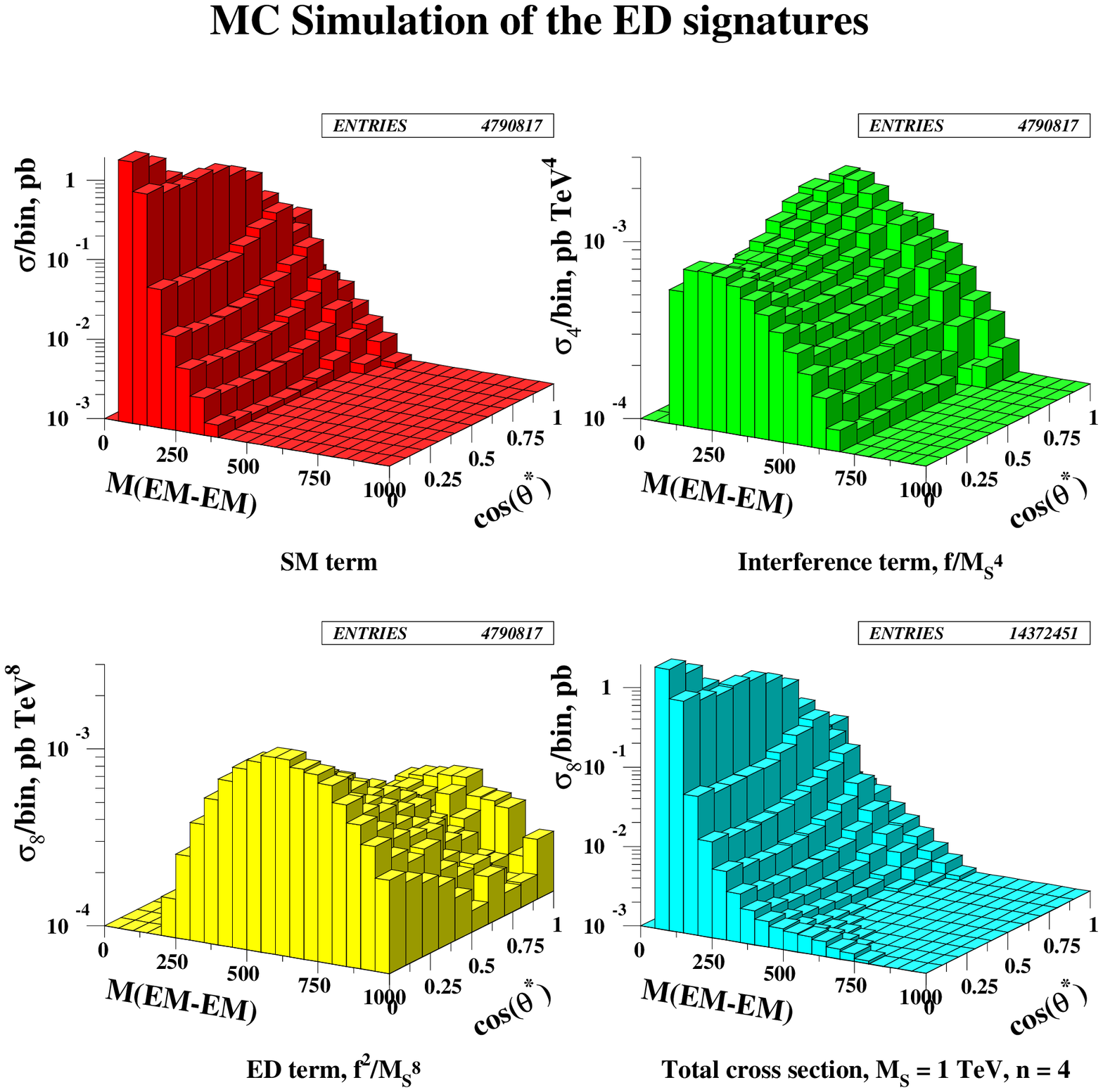}
 \end{minipage}\hfill
\renewcommand{\baselinestretch}{1.0}
 \caption{\it a) Comparison in the $M_{EM-EM}$ vs $|cos(\theta^{*}|$ plane
 of Data with Monte Carlo for SM processes ($EM=$ $\gamma$ or $e$);
b) same as in a) including also the $LED$ contributes.}
 \label{D0fig}
 \end{figure}

\noindent energy is above the Planck Scale ($\sqrt{s} > M_{Pl}$).
In $LED$ models the Planck Scale can be effectively
${\mathcal{O}}(1\,TeV)$, opening up the interesting possibility of
producing and studying $BH$ using collider
experiments~\cite{BH-1}. The observability of such $BH$ at future
colliders will depend on the value of the fundamental Planck
Scale. At Hadron colliders, where it is possible to reach the
highest center of mass energies, compared to other machines,
the $BH$ production cross section can be written as follow:
\begin{equation}
\sigma_{pp \rightarrow BH}(M_{min},s) =
\sum_{ij} \int^{1}_{\tau_{m}} d\tau \int^{1}_{\tau} \frac{dx}{x}
F_{i}(x) F_{j}(\tau/x)\sigma_{ij \rightarrow BH} (\tau s)
\end{equation}
where $i$ and $j$ are the two colliding partons,
$x$ and $\tau/x$ are the momentum fractions of $i$ and $j$
and $F$ are the parton distribution functions.
If we assume the Thorne's hoop conjecture, that states that horizons
form when and only when a mass M is compacted into a region whose
circumference in every directions is less than $2 \pi R_{BH}(M)$, we
obtain the important result that the black hole  production
cross section is :
\begin{equation}
\sigma_{ij \rightarrow BH}(s) \sim \pi R^{2}_{BH}(\sqrt{s})
\end{equation}
The precise mass of the $BH$ formed in a collision depends on the
amount of energy and matter which becomes trapped behind the
event horizon. If the scale of gravity is the $TeV$ scale, $BH$
production could be a dominant process at hadron colliders
\newpage
\begin{table}[t!]
\begin{tabular}{c|@{}cccccc|@{}cc}
\hline\hline
GRW~\cite{GRW} & \multicolumn{6}{@{}c|}{HLZ~\cite{HLZ}} & \multicolumn{2}{@{}c}{Hewett~\cite{Hewett}} \\
\hline\hline
& ~~$n$=2 & $n$=3 & $n$=4 & $n$=5 & $n$=6 & $n$=7~~ & ~~$\lambda=+1$ & $\lambda=-1$ \\
\cline{2-7} \cline{8-9}
1.2 & ~~1.4 & 1.4     & 1.2   & 1.1   & 1.0   & 1.0~~  & ~~1.1      & 1.0 \\
\hline\hline
\end{tabular}
\caption{Lower limits at 95\% CL on the effective Planck Scale, $M_S$, in $TeV$.}
\label{D0tab}
\end{table}
\noindent
beyond
LHC. For example for $M_{Pl} = 1$ $TeV$ and ${\mathcal{D}}= 10$, at
the Very Large Hadron Collider (VLHC) considering $\sqrt{s}=$ $100$
$TeV$ (something in the middle between the stage I and the stage II)
assuming an integrated luminosity per year of $100$ $fb^{-1}yr^{-1}$, $BH$ of
mass around 10 $TeV$ should be produced with a rate of $1$ $Kz$.

\subsection{Black Hole decays and signatures}

Once produced black holes decay. The decay process is rather complex
and occurs in several stages~\cite{BH-2}:

\begin{enumerate}
\item {\it Balding Phase}: in which there is emission of gauge and
gravitational radiation that will settle down the $BH$ to a symmetrical
rotating object with a growing horizon;
\item {\it Spin-down Phase}: in which the $BH$ Hawking radiates,
emitting quanta having angular moment $\ell \sim 1$;
\item {\it Schwarzschild-Hawking Evaporation Phase}:
The Spin-down phase leaves a Schwarzschild $BH$ that continues to radiate
Hawking Radiation; instant quanta are emitted with a thermal spectrum
around Hawking temperature ($T_{H}$);
\item {\it Planck Phase}
Once the $BH$ reach the Planck Mass $M \sim M_{p}$ the $BH$ completely
decays emitting few quanta with energies ${\mathcal{O}}(M_{p})$;
\end{enumerate}
Because of the large cross section, multiplicity and visible
energy, $BH$, at hadron colliders, give rise to very spectacular
events. The $BH$ production and decay is characterized by the
following specific signatures~\cite{Greg,BH-3}:
\begin{itemize}

\item suppression of hard perturbative scattering processes
at energy in which the $BH$ production start to dominate;

\item very large production cross sections that grows with the
energy;

\item high multiplicity events
(for $M_{BH} \sim 10$ $TeV$, ${\mathcal{D}}=10$ the expected
multiplicity is $\sim 50$) with visible transverse energy of the
order of $\sim 1/3$ of the total energy;

\item high sphericity events because of the small $BH$ boost
($\langle \gamma\beta \rangle$ $<$ $1$);

\item a ratio of $\sim 1/5$ between leptonic and hadronic
activity because of the Schwarzschild-Hawking Evaporation Phase.

\end{itemize}


\newpage
\begin{table}\hspace{.0cm}
\medskip
\centering
\hspace{-1.6cm}
\begin{tabular}{|c|c|c|c|}
\hline\hline
\multicolumn{4}{|c|}{(a) Hadron Colliders} \\
\hline
\hline
Run I & Run IIa & Run IIb & LHC  \\
${\cal L}=0.13$ fb$^{-1}$ & 2 & 20 & 100 \\
\hline
1.3 & 1.9 & 2.6 & 9.9 \\
\hline\hline
\end{tabular}
\hspace{-0.2cm}
\begin{tabular}{|cc|cc|cc|}
\hline\hline
\multicolumn{6}{|c|}{(b) $e^+ e^-$ Colliders} \\
\hline
\hline
\multicolumn{2}{|c|}{$\sqrt{s}=0.5$ TeV} &
\multicolumn{2}{c|}{1 TeV } & \multicolumn{2}{c|}{1.5 TeV } \\
${\cal L}=10$ fb$^{-1}$     & 50  & 50  & 100 & 100  & 200 \\
\hline
3.9 & 4.8 & 7.9 & 8.9 & 12.0 & 13.0   \\
\hline\hline
\end{tabular}
\renewcommand{\baselinestretch}{1.0}
\hspace{-1.6cm}
\caption{95\% C.L. sensitivity limits on $M_S\;(n=4)$ at (a) hadronic colliders
(Tevatron and LHC) and (b) $e^+ e^-$ colliders of 0.5 -- 1.5 TeV.}
 \label{future}
\end{table}


\section{Black Hole production and Cosmic Rays}

There are other possibilities to study experimentally the Black
Hole production associated with the existence of extra spatial
dimensions. Black Holes, in fact, may be produced by inelastic
scattering of ultra-high energy cosmic neutrinos with the matter.
As Black Hole decays give rise to deeply penetrating showers, with
an electromagnetic component which differs substantially from the
conventional neutrino interactions~\cite{Anchordoqui:2001ei}, it
is possible to reach good signal on background discrimination. At
the present BH production can be studied using cosmic rays in the
interaction with
air nuclei and detected by air shower arrays like in the case of
the Auger Experiment~\cite{Feng:2001ib,Ringwald:2001vk},
or in the interaction with ice or water nuclei like in the case of
neutrino telescopes such as AMANDA,
IceCube~\cite{Uehara:2001yk,OTH} and RICE~\cite{Frichter:1999kr}.
These cosmic ray opportunities can be realized before the start of
the physics run of the Large Hadron Collider.

\vspace{-0.7cm}
\section{Acknowledgments}

The author wants to thank Laszlo Jenkovszky and the Organizers of
the International Conference: New Trends in High Energy Physics
for the kind invitation to deliver this lecture. I would like also
to thank G. Landsberg and A. Ringwald for their suggestions and
comments.



\eject
\end{document}